\documentclass[11pt]{article}
\usepackage{latexsym}
\usepackage{graphicx}

\def\AFOUR{%
\setlength{\textheight}{8.5in}%
\setlength{\textwidth}{5.75in}%
\setlength{\topmargin}{-0.375in}%
\hoffset=-.5in%
\renewcommand{\baselinestretch}{1.17}%
\setlength{\parskip}{6pt plus 2pt}%
}

\AFOUR                                           

\expandafter\ifx\csname amssym.def\endcsname\relax \else\endinput\fi
\expandafter\edef\csname amssym.def\endcsname{%
       \catcode`\noexpand\@=\the\catcode`\@\space}
\catcode`\@=11
\def\undefine#1{\let#1\undefined}
\def\newsymbol#1#2#3#4#5{\let\next@\relax
 \ifnum#2=\@ne\let\next@\msafam@\else
 \ifnum#2=\tw@\let\next@\msbfam@\fi\fi
 \mathchardef#1="#3\next@#4#5}
\def\mathhexbox@#1#2#3{\relax
 \ifmmode\mathpalette{}{\m@th\mathchar"#1#2#3}%
 \else\leavevmode\hbox{$\m@th\mathchar"#1#2#3$}\fi}
\def\hexnumber@#1{\ifcase#1 0\or 1\or 2\or 3\or 4\or 5\or 6\or 7\or 8\or
 9\or A\or B\or C\or D\or E\or F\fi}

\font\tenmsa=msam10
\font\sevenmsa=msam7
\font\fivemsa=msam5
\newfam\msafam
\textfont\msafam=\tenmsa
\scriptfont\msafam=\sevenmsa
\scriptscriptfont\msafam=\fivemsa
\edef\msafam@{\hexnumber@\msafam}
\mathchardef\dabar@"0\msafam@39
\def\dashrightarrow{\mathrel{\dabar@\dabar@\mathchar"0\msafam@4B}}
\def\dashleftarrow{\mathrel{\mathchar"0\msafam@4C\dabar@\dabar@}}

\def\ulcorner{\delimiter"4\msafam@70\msafam@70 }
\def\urcorner{\delimiter"5\msafam@71\msafam@71 }
\def\llcorner{\delimiter"4\msafam@78\msafam@78 }
\def\lrcorner{\delimiter"5\msafam@79\msafam@79 }
\def\yen{{\mathhexbox@\msafam@55}}
\def\checkmark{{\mathhexbox@\msafam@58}}
\def\circledR{{\mathhexbox@\msafam@72}}
\def\maltese{{\mathhexbox@\msafam@7A}}
\def\circledS{{\mathhexbox@\msafam@73}}

\font\tenmsb=msbm10
\font\sevenmsb=msbm7
\font\fivemsb=msbm5
\newfam\msbfam
\textfont\msbfam=\tenmsb
\scriptfont\msbfam=\sevenmsb
\scriptscriptfont\msbfam=\fivemsb
\edef\msbfam@{\hexnumber@\msbfam}
\def\Bbb#1{{\fam\msbfam\relax#1}}
\def\widehat#1{\setbox\z@\hbox{$\m@th#1$}%
 \ifdim\wd\z@>\tw@ em\mathaccent"0\msbfam@5B{#1}%
 \else\mathaccent"0362{#1}\fi}
\def\widetilde#1{\setbox\z@\hbox{$\m@th#1$}%
 \ifdim\wd\z@>\tw@ em\mathaccent"0\msbfam@5D{#1}%
 \else\mathaccent"0365{#1}\fi}
\font\teneufm=eufm10
\font\seveneufm=eufm7
\font\fiveeufm=eufm5
\newfam\eufmfam
\textfont\eufmfam=\teneufm
\scriptfont\eufmfam=\seveneufm
\scriptscriptfont\eufmfam=\fiveeufm
\def\frak#1{{\fam\eufmfam\relax#1}}

\csname amssym.def\endcsname

\makeatletter
\def\section{\@startsection{section}{1}{\z@}{-3.5ex plus -1ex minus
 -.2ex}{2.3ex plus .2ex}{\large\sc}}
\def\subsection{\@startsection{subsection}{2}{\z@}{-3.25ex plus -1ex minus
 -.2ex}{1.5ex plus .2ex}{\normalsize\sc}}
\makeatother

\makeatletter
\@addtoreset{equation}{section}

\makeatother

\newcommand{\nc}{\newcommand}
\newcommand{\rnc}{\renewcommand}

\nc{\subs}[1]{{\vspace*{0.5cm}}%
{\noindent\underline{\small\sc #1}}{\addcontentsline{toc}{subsubsection}{#1}}%
{\vspace*{0.3cm}}}

\nc{\subss}[1]{{\vspace*{0.5cm}}%
{\noindent\underline{\small\sc #1}}%
{\vspace*{0.3cm}}}

\nc{\chap}[1]{{\clearpage}%
\begin{center}%
{\noindent\underline{\large\sc #1}}{\addcontentsline{toc}{section}{#1}}%
\end{center}%
{\vspace*{0.3cm}}}

\nc{\be}{\begin{equation}}
\nc{\ee}{\end{equation}}
\nc{\bea}{\begin{eqnarray}}
\nc{\eea}{\end{eqnarray}}

\nc{\trac}[2]{{\textstyle\frac{#1}{#2}}}

\nc{\ex}[1]{\mbox{e}^{\,\textstyle#1}}

\nc{\CC}{\Bbb{C}}
\nc{\HH}{\Bbb{H}}
\nc{\PP}{\Bbb{P}}
\nc{\RR}{\Bbb{R}}
\nc{\ZZ}{\Bbb{Z}}
\nc{\II}{\Bbb{I}}
\nc{\EE}{\Bbb{E}}

\rnc{\a}{\alpha}
\rnc{\b}{\beta}
\rnc{\d}{\delta}
\nc{\ga}{\gamma}
\nc{\la}{\lambda}
\nc{\f}{\phi}
\nc{\p}{\psi}
\nc{\e}{\eta}
\rnc{\c}{\chi}
\nc{\eps}{\epsilon}
\nc{\om}{\omega}
\nc{\Om}{\Omega}

\nc{\symx}{\circledS}
\newsymbol\smallsmile 1360
\newsymbol\smallfrown 1361
\nc{\ad}{\mathop{\mbox{ad}}\nolimits}
\nc{\tr}{\mathop{\mbox{tr}}\nolimits}
\nc{\Tr}{\mathop{\mbox{Tr}}\nolimits}
\nc{\Det}{\mathop{\mbox{Det}}\nolimits}
\rnc{\det}{\mathop{\mbox{det}}\nolimits}
\nc{\rk}{\mathop{\mbox{rk}}\nolimits}
\nc{\del}{\partial}
\nc{\diag}{\mathop{\mbox{diag}}\nolimits}
\nc{\ra}{\rightarrow}
\nc{\Ra}{\Rightarrow}
\nc{\LRa}{\Leftrightarrow}
\nc{\lra}{\leftrightarrow}
\nc{\ot}{\otimes}
\rnc{\ss}{\subset}
\nc{\nul}{\noindent\underline}
\nc{\non}{\nonumber\\}
\nc{\mat}[4]{\left(\begin{array}{cc}#1&#2\\#3&#4\end{array}\right)}
\rnc{\lg}{\frak{g}}
\nc{\G}[3]{\Gamma^{#1}_{\;{#2}{#3}}}
\nc{\nam}{\nabla_{\mu}}
\nc{\nan}{\nabla_{\nu}}
\nc{\dx}{\dot{x}}
\nc{\dxl}{\dot{x}^{\la}}
\nc{\dxm}{\dot{x}^{\mu}}
\nc{\dxn}{\dot{x}^{\nu}}
\nc{\ddx}{\ddot{x}}
\nc{\ddxm}{\ddot{x}^{\mu}}
\nc{\ddxn}{\ddot{x}^{\nu}}
\nc{\dxi}{\dot{\xi}}
\nc{\ddxi}{\ddot{\xi}}
\nc{\xp}{x^{+}}
\nc{\xm}{x^{-}}

\begin{document}
\thispagestyle{empty}

\vspace*{1cm}
\begin{center}
{\Large\sc Fermi Coordinates and Penrose Limits}
\end{center}
\vspace{0.7cm}

\begin{center}
{\large\textsc{Matthias Blau, Denis Frank, Sebastian Weiss}}
\end{center}

\centerline{\it Institut de Physique, Universit\'e de Neuch\^atel}
\centerline{\it Rue Breguet 1, CH-2000 Neuch\^atel, Switzerland}

\vfill
\begin{center}
\textbf{Abstract}
\end{center}

We propose a formulation of the Penrose plane wave limit in terms of
null Fermi coordinates. This provides a physically intuitive (Fermi
coordinates are direct measures of geodesic distance in space-time) and
manifestly covariant description of the expansion around the plane wave
metric in terms of components of the curvature tensor of the original
metric, and generalises the covariant description of the lowest order
Penrose limit metric itself, obtained in \cite{bbop1}. We
describe in some detail the construction of null Fermi coordinates and
the corresponding expansion of the metric, and then study various aspects
of the higher order corrections to the Penrose limit. In particular,
we observe that in general the first-order corrected metric is such 
that it admits a light-cone gauge description in string theory.
We also establish a formal analogue of the Weyl tensor peeling theorem for
the Penrose limit expansion in any dimension, and we give a simple
derivation of the leading (quadratic) corrections to the Penrose limit
of $\mathrm{AdS}_5\times S^5$.

\vfill
\newpage

\begin{scriptsize}
\tableofcontents
\end{scriptsize}


\section{\sc Introduction}

Following the observations in \cite{bfhp1,rrm,bfhp2,bmn} regarding the
maximally supersymmetric type IIB plane wave background, its relation to
the Penrose limit of $\mathrm{AdS}_5 \times S^5$, and the corresponding
BMN limit on the dual CFT side\footnote{see e.g.\ \cite{sheikh} for a review
and further references.}, the Penrose plane wave limit construction
\cite{penrose} has attracted a lot of attention.
This construction associates to a Lorentzian
space-time
metric $g_{\mu\nu}$ and a null-geodesic $\gamma$ in that space-time a plane
wave metric, 
\be
(ds^2=g_{\mu\nu}dx^{\mu}dx^{\nu},\gamma) \;\;\;\;\ra\;\;\;\;
d\bar{s}_{\gamma}^2 = 2 d\xp d\xm + A_{ab}(\xp)x^ax^b d\xp{}^2 + 
\d_{ab}dx^a dx^b\;\;,
\label{pl1}
\ee
the right hand side being the metric of a plane wave in Brinkmann coordinates,
characterised by the wave profile $A_{ab}(\xp)$.

The usual definition of the Penrose limit \cite{penrose,gueven,bfp} is
somewhat round-about and in general requires a sequence of coordinate
transformations (to adapted or Penrose coordinates, from Rosen
to Brinkmann coordinates), scalings (of the metric and the adapted
coordinates) and limits.\footnote{For sufficiently simple metrics and
null geodesics it is of course possible to devise more direct ad hoc
prescriptions for finding a Penrose limit.} And even though general
arguments about the covariance of the Penrose limit \cite{bfp} show that
there is of course something covariant lurking behind that prescription,
after having gone through this sequence of operations one has probably
pretty much lost track of what sort of information about the original
space-time the Penrose limit plane wave metric actually encodes.

This somewhat unsatisfactory state of affairs was improved upon
in \cite{bbop1,bbop2}. There it was shown that the wave profile
$A_{ab}(x^+)$ of the Penrose limit metric can be determined from the
original metric without taking any limits, and has a manifestly covariant
characterisation as the matrix
\be
A_{ab}(x^+) = -R_{a+b+}|_{\gamma(x^+)}
\label{imain}
\ee
of curvature components (with respect to a suitable frame) of the original
metric, restricted to the null geodesic $\gamma$. This will be briefly
reviewed in section 2.

The aim of the present paper is to extend this to a covariant prescription 
for the expansion of the original metric around the Penrose limit metric, 
i.e.\ to find a formulation of the Penrose limit which is such that
\begin{itemize}
\item to lowest order one directly finds the plane wave metric in Brinkmann
coordinates, with the manifest identification (\ref{imain});
\item higher order corrections are also covariantly expressed in terms of
the curvature tensor of the original metric.
\end{itemize}

We are thus seeking analogues of Brinkmann coordinates, the covariant
counterpart of Rosen coordinates for plane waves, for an arbitrary
metric. We will show that this is provided by Fermi coordinates
based on the null geodesic $\gamma$. Fermi normal coordinates for
\textit{timelike} geodesics are well known and are discussed in detail
e.g.\ in \cite{synge,poisson}. They are natural coordinates for freely falling
observers since, in particular, the corresponding Christoffel symbols
vanish along the entire worldline of the observer (geodesic), thus embodying
the equivalence principle.

In retrospect, the appearance of Fermi coordinates in this context is
perhaps not particularly surprising. Indeed, it has always been clear
that, in some suitable sense, the Penrose limit should be thought of as a
truncation of a Taylor expansion of the metric in directions transverse to
the null geodesic, and that the full expansion of the metric should just
be the complete transverse expansion. The natural setting for a covariant
transverse Taylor expansion are Fermi coordinates, and thus what we are
claiming is that the precise way of saying ``in some suitable sense''
is ``in Fermi coordinates''.

In order to motivate this and to understand how to generalise
Brinkmann coordinates, in section 3 we will begin with some elementary
considerations, showing that Brinkmann coordinates are null Fermi
coordinates for plane waves. Discussing plane waves from this point
of view, we will also recover some well known facts about Brinkmann
coordinates from a slightly different perspective.

In section 4 we introduce null Fermi coordinates in general, adapting
the construction  of timelike Fermi coordinates in \cite{poisson} to the
null case. These coordinates $(x^A)=(x^+,x^{\bar{a}})$ consist of the affine
parameter $x^+$ along the null geodesic $\gamma$ and geodesic coordinates
$x^{\bar{a}}$ in the transverse directions. 
We also introduce the covariant transverse Taylor
expansion of a function, which takes the form
\be
f(x) = \sum_{n=0}^{\infty}\frac{1}{n!}
\left(E^{\mu_1}_{\bar{a}_1}\ldots E^{\mu_n}_{\bar{a}_n}
\nabla_{\mu_1}\ldots\nabla_{\mu_n}f\right)(x^+)\;x^{\bar{a}_1}\ldots
 x^{\bar{a}_n} \;\;,
\label{itaylor2}
\ee
where $E^{\mu}_A$ is a parallel frame along $\gamma$.
As an application we
show that the coordinate transformation
from arbitrary adapted coordinates (i.e.\ coordinates for which the null
geodesic $\gamma$ agrees with one of the coordinate lines) to Fermi
coordinates is nothing other than the transverse Taylor expansion of
the coordinate functions in terms of Fermi coordinates.

In section 5, we discuss the covariant expansion of the metric in
Fermi coordinates in terms of components of the Riemann tensor and
its covariant derivatives evaluated on the null geodesic. We explicitly
derive the expansion of the metric up to quadratic order in the transverse
coordinates and show that the result is the exact null analogue of the
classical Manasse-Misner result \cite{mami} in the timelike case, namely
\bea
ds^2&=& 2dx^+dx^- +\delta_{ab} dx^a dx^b \non
&-&\left[R_{+\bar{a}+{\bar{b}}} \ x^{\bar{a}} x^{\bar{b}} (dx^+)^2 
         +\trac43 R_{+{\bar{b}}\bar{a}{\bar{c}}} 
          x^{\bar{b}} x^{\bar{c}} (dx^+ dx^{\bar{a}}) +\trac13
          R_{\bar{a}{\bar{c}}{\bar{b}}{\bar{d}}}  x^{\bar{c}} x^{\bar{d}}
          (dx^{\bar{a}} dx^{\bar{b}})\right]\non
   &+& \mathcal{O}(x^{\bar{a}}x^{\bar{b}}x^{\bar{c}})
\label{imetex}
\eea
where $(x^{\bar{a}})=(x^-,x^a)$ and
all the curvature components are evaluated on $\gamma$.
The expansion up to quartic order in the transverse coordinates is given
in appendix A.1.

In section 6, we show how to implement the Penrose limit in Fermi
coordinates. To that end we first discuss the behaviour of Fermi
coordinates under scalings $g_{\mu\nu}\ra\la^{-2}g_{\mu\nu}$ of the
metric. Since Fermi coordinates are geodesic coordinates, measuring
invariant geodesic distances, Fermi coordinates will scale non-trivially
under scalings of the metric, and we will see that the characteristic
asymmetric scaling of the coordinates that one performs in whichever way
one does the Penrose limit arises completely naturally from the very
definition of Fermi coordinates. Combining this with the expansion of
the metric of section 5, we then obtain the desired covariant expansion
of the metric around its Penrose limit.

The expansion to $\mathcal{O}(\lambda)$, for which knowledge of the
expansion of the metric in Fermi coordinates to cubic order is required,
reads
\bea
ds^2 &=& 2dx^+dx^- + \d_{ab}dx^a dx^b -R_{a+b+}x^ax^b (dx^+)^2\non
     &+& \lambda\left[-2 R_{+a+-} \ x^a x^-
(dx^+)^2 -\trac43 R_{+bac} \ x^b x^c (dx^+ dx^a) 
-\trac13 R_{+a+b;c} \ x^a x^b x^c (dx^+)^2 \right]\non
&+& \mathcal{O}(\lambda^2)\;\;.
\eea
where the first line is the Penrose limit metric (\ref{pl1}). 
In particular, if the characteristic covariantly
constant null vector $\del/\del x^-$ of (\ref{pl1}) is such that it
remains Killing at first order it is actually covariantly constant and
the first-order corrected metric is that of a pp-wave which is amenable
to a standard light-cone gauge description in string theory \cite{hs}.
Moreover, in general the above metric is precisely such that it admits
a modified light-cone gauge in the sense of \cite{rudd}.
The expansion to $\mathcal{O}(\la^2)$ is given in appendix A.2. 


We illustrate the formalism in section 7 by giving a quick
derivation of the second order corrections to the Penrose limit of
$\mathrm{AdS}_{5}\times S^5$. These corrections have been calculated
before in other ways \cite{ryzhov,callan}, and the point of this example
is not so much to advocate the Fermi coordinate prescription as the
method of choice to do such calculations (even though it is geometrically
appealing and transparent in general, and the calculation happens to be
extremely simple and purely algebraic in this particular case).  Rather,
the interest is more conceptual and lies in the precise
identification of the
corrections that have already been calculated (and subsequently been
used in the context of the BMN correspondence) with particular
components of the curvature tensor of $\mathrm{AdS}_{5}\times S^5$.

In section 8 we return to the general structure of the $\lambda$-expansion
of the metric. The leading non-trivial contribution to the metric is
the $\lambda^0$-term $R_{a+b+}$ (\ref{imain}) of the Penrose limit, and
higher order corrections involve other frame components of the Riemann
tensor, each arising with a particular scaling weight $\lambda^w$. In
the four-dimensional case it was shown in \cite{kunze}, using the
Newman-Penrose formalism, that the complex Weyl scalars $\Psi_i$,
$i=0,\ldots,4$ scale as $\la^{4-i}$.  This is formally analogous to the
scaling of the $\Psi_i$ as $(1/r)^{5-i}$ with the radial distance, the
peeling theorem \cite{peeling} of radiation theory in general relativity.
We will show that the present covariant formulation of the Penrose limit
significantly simplifies the analysis of the peeling property in this
context (already in dimension four) and, using the analysis in \cite{cmpp}
of algebraically special tensors and the (partial) generalised Petrov
classification of the Weyl tensor in higher dimensions, allows us to
establish an analogous result in any dimension.

We hope that the covariant null Fermi normal coordinate expansion
of the metric developed here will provide a useful alternative to the
standard Riemann normal coordinate expansion, in particular, but not
only, in the context of string theory in plane wave backgrounds and
perturbations around such backgrounds.

\section{\sc Lightning Review of the Penrose Limit}

The traditional systematic construction of the Penrose limit
\cite{penrose,gueven,bfp} involves the following steps:
\begin{enumerate}
\item First one introduces Penrose coordinates 
$(U,V,Y^k)$ adapted to the null geodesic $\gamma$
(see \cite{bbop2} for a general construction),
in which the metric takes the form
\be
ds_{\gamma}^2 =
2 dUdV + a(U,V,Y^k) dV^2 + 2 b_i(U,V,Y^k) dY^i dV + g_{ij}(U,V,Y^k)
dY^i dY^j\;\;.
\label{acs2}
\ee
Here the original null-geodesic $\gamma$ 
is the curve $(U,0,0)$ with affine parameter
$U$, embedded into the congruence $(U,V_0,Y^i_0)$ of null geodesics labelled
by the constant values $(V_0,Y^i_0)$, $i=1,\ldots,d$,
of the transverse coordinates.
\item
Next one performs an asymmetric rescaling of the coordinates,
\be
(U,V,Y^{k})=(u,\lambda^2 v,\lambda y_{k}) \;\;,
\label{dsdsl}
\ee
accompanied by an overall rescaling of the metric, to obtain
the one-parameter family of metrics
\bea
\la^{-2}ds_{\gamma,\lambda}^2 &=&
2 dudv + \la^2 a(u,\la^2v,\la y^k) dv^2 +
2 \la b_i(u,\la^2v,\la y^k) dy^i dv\non
&& + g_{ij}(u,\la^2v,\la y^k) dy^i dy^j\;\;.
\label{ldsgl}
\eea
\item
Now taking the
combined infinite boost and large volume limit $\la\ra 0$ results in a
well-defined and non-degenerate metric $\bar{g}_{\mu\nu}$,
\bea
\mathrm{Penrose\;Limit:}\;\;
d\bar{s}_{\gamma}^2
&=& \lim_{\lambda\ra 0}\lambda^{-2}ds_{\gamma,\lambda}^2\label{limit}\\
&=& 2dudv + \bar{g}_{ij}(u)dy^i dy^j\;\;,
\label{rcs}
\eea
where $\bar{g}_{ij}(u)= g_{ij}(u,0,0)$ is the restriction of
$g_{ij}$ to the null geodesic $\gamma$. This is the metric
of a plane wave in Rosen coordinates.
\item 
One then transforms this to Brinkmann coordinates $(x^A)=(\xp,\xm,x^a)$, 
$a=1,\ldots,d$, via
\be
(u,v,y^k) = 
(\xp, \xm + \trac{1}{2}\dot{\bar{E}}_{ai}\bar{E}^i_{\;b}x^a x^b,
\bar{E}^k_{\;a}x^a)
\label{rcbc}
\ee
where $\bar{E}^a_{\;i}$ is a vielbein for $\bar{g}_{ij}$, i.e.\
$\bar{g}_{ij}=\bar{E}^a_{\;i}\bar{E}^b_{\;j}\d_{ab}$,
required to satisfy the symmetry condition
$\dot{\bar{E}}_{ai}\bar{E}^i_{\;b} = \dot{\bar{E}}_{bi}\bar{E}^i_{\;a}$.
In these coordinates 
the plane wave metric takes the canonical form
\be
d\bar{s}_{\gamma}^2 = 2 d\xp d\xm + A_{ab}(\xp)x^ax^b d\xp{}^2 + \d_{ab}dx^a
dx^b\;\;,
\label{bc}
\ee
with $A_{ab}(\xp)$ given by \cite{mmhom}
\be
A_{ab}= \ddot{\bar{E}}_{ai} \bar{E}^i_{\;b}\;\;.
\ee
\end{enumerate}

While this is, in a nutshell, the construction of the Penrose limit
metric, the above definition looks rather round-about and non-covariant
and manages to hide quite effectively the relation between the original
data $(g_{\mu\nu}, \gamma)$ and the resulting plane wave metric.
In principle taking the Penrose limit amounts
to assigning the wave profile $A_{ab}$ to the initial data
$(g_{\mu\nu},\gamma)$, 
\be
(g_{\mu\nu},\gamma)\;\;\;\;\ra\;\;\;\;A_{ab}\;\;.
\ee
This certainly begs
the question if there is not a more direct (and geometrically appealing)
route from $(g_{\mu\nu},\gamma)$ to $A_{ab}$ which elucidates the
precise nature of the Penrose limit and the extent to which it
encodes generally covariant properties of the original space-time.

Indeed, as shown in \cite{bbop1,bbop2}, there is. Given the affinely
parametrised null geodesic $\gamma=\gamma(u)$, the tangent vector
$E_+^{\mu}=\dot{\gamma}^{\mu}$ is (by definition) parallel transported 
along $\gamma$. We extend this to a pseudo-orthonormal parallel transported
frame $(E_A^{\mu})=(E_+^{\mu},E_-^{\mu},E_a^{\mu})$ along $\gamma$. 
Thus, in terms of the dual coframe $(E^A_{\mu})$, the metric restricted to 
$\gamma$ can be written as
\be
ds^2|_{\gamma} = 2E^+E^- + \d_{ab}E^aE^b\;\;.
\ee
The main result of \cite{bbop1} is the observation that the wave profile
$A_{ab}(x^+)$ 
of the associated Penrose limit metric is  nothing other than the matrix
\be
A_{ab}(x^+) = -R_{a+b+}|_{\gamma(x^+)}
\label{main}
\ee
of frame curvature components of the original metric, evaluated 
at the point $\gamma(x^+)$.

Modulo constant $SO(d)$-rotations this is independent of the choice of 
parallel frame and provides 
a manifestly covariant characterisation of the Penrose limit
plane wave metric which, moreover, does not require taking any limits.
The geometric significance of $A_{ab}(x^+)$ is that it is the
transverse null geodesic deviation matrix along $\gamma$ 
\cite[Section 4.2]{HE} of the original metric, 
\be 
\frac{d^2}{du^2} Z^a = A_{ab}(u)
Z^b\;\;, \label{gde} 
\ee 
with $Z$ the transverse geodesic deviation vector. 
Since the only non-vanishing curvature components of the Penrose limit
plane wave metric
$d\bar{s}_{\gamma}^2$ in Brinkmann coordinates (\ref{bc}) are 
\be
\bar{R}_{a+b+} = -A_{ab}\;\;,
\label{rbar}
\ee
this implies that geodesic deviation along the selected null
geodesic in the original space-time is identical to null geodesic
deviation in the corresponding Penrose limit plane wave metric
and shows that it is precisely this information about tidal forces in 
the original metric that the Penrose limit encodes (while discarding 
all other information about the original metric).

Let us now consider higher order terms in the expansion of the original metric
about the Penrose limit. To that end we return to (\ref{ldsgl}) and expand
in a power-series in $\lambda$. To $\mathcal{O}(\lambda)$ one has
\bea
\la^{-2}ds_{\gamma,\lambda}^2 &=&
2 dudv + \bar{g}_{ij}(u)dy^i dy^j \non
 &+&  \lambda\left(2 \bar{b}_i(u) dy^i dv +
y^k\bar{g}_{ij,k}(u)dy^idy^j\right) + \mathcal{O}(\lambda^2)
\eea
where, as before, an overbar denotes evaluation on the null geodesic, i.e.\
$\bar{g}_{ij,k}(u)= g_{ij,k}(u,0,0)$ etc. We see that in the expansion
of the metric in Penrose coordinates these higher order 
terms are not covariant (e.g.\ the $\bar{g}_{ij,k}$ are Christoffel symbols). 

This raises the question if there is a different way of implementing the
Penrose limit which is such that all terms in the $\lambda$-expansion
of the metric are covariant expressions in the curvature tensor of the
original metric.


A ham-handed way to approach this issue would be to seek a
$\lambda$-dependent (and analytic in $\lambda$) coordinate transformations
that extends the transformation from Rosen to Brinkmann coordinates
and, applied to the above expansion of the metric, results in order by
order covariant expressions. However, first of all this strategy puts
undue emphasis on the coordinate transformation that relates Penrose
coordinates to the new coordinates, rather than on the expansion of
the metric itself. Secondly, even if one happens to find a solution to
the problem in this way, in all likelihood one will in the end have
discovered a coordinate system that is sufficiently natural to have
been discoverable by other, less brute-force, means as well. Indeed, we
will see in sections 5 and 6, without having to go through the explicit
coordinate transformation from Penrose coordinates, that all this is
accomplished by Fermi coordinates adapted to the null geodesic $\gamma$.

\section{\sc Brinkmann Coordinates are Null Fermi Coordinates}

In this section we will discuss Brinkmann coordinates for plane waves from
(what will turn out to be) the point of view of Fermi coordinates. The
considerations in this section are elementary, but they serve as a
motivation for the subsequent general discussion of Fermi coordinates.
Moreover, we find it illuminating to recover some well known facts
about Brinkmann coordinates and their relation to Rosen coordinates from
this perspective.

First of all, we note that a particular solution of the null geodesic
equation in Brinkmann coordinates is the curve $\gamma(u)= (u,0,0)$
with affine parameter $u = x^+$ (in the Penrose limit context this is
obviously just the original null geodesic $\gamma$).  Along this curve all
the Christoffel symbols of the metric are zero (the a priori non-vanishing
Christoffel symbols are linear and quadratic in the $x^a$ and thus vanish
for $x^a=0$). This is the counterpart of the usual statement for Riemann
normal coordinates that the Christoffel symbols are zero at some chosen
base-point. Here we have a geodesic of such base-points.

Next we observe that the straight lines 
\be
x^A(s) = (x^+_0,sx^-,sx^a) 
\ee
connecting a point $(x^+_0,0,0)$ on $\gamma$ to the point $(x^+_0,x^-,x^a)$
are also geodesics. In the standard plane wave terminology these are
spacelike or null geodesics with zero lightcone momentum, $p_-=
x^{+\prime}(s)=0$, a  prime denoting an $s$-derivative. Thus the coordinate 
lines of $x^-$ and $x^a$ are geodesics, while $x^+$ labels the original
null geodesic $\gamma$.
These are the characteristic and defining properties of null Fermi coordinates.

There is also a Fermi analogue of the Riemann normal coordinate expansion of
the metric in terms of the Riemann tensor and its covariant derivatives. In
the special case of plane waves we have, combining (\ref{bc}) with
(\ref{rbar}), 
\be
d\bar{s}^2 = 2dx^+dx^- + \d_{ab}dx^a dx^b -
\bar{R}_{a+b+}(x^+)x^ax^b dx^+{}^2\;\;.
\label{bc2}
\ee
Thus in this case the expansion of the metric terminates at quadratic order.

We can also understand (and rederive) the somewhat peculiar coordinate
transformation (\ref{rcbc}) from Rosen to Brinkmann coordinates from
this point of view. Thus this time we begin with the metric 
\be
d\bar{s}^2 = 2dudv + \bar{g}_{ij}(u)dy^idy^j
\ee
of a plane wave in Rosen coordinates and introduce a pseudo-orthonormal 
frame $\bar{E}^{\mu}_A$, 
\be
\bar{E}_+=\del_u\;\;,\;\;\bar{E}_-=\del_v\;\;,\;\;\bar{E}_a=\bar{E}_a^i\del_i
\label{rcpof}
\ee
where $\bar{E}^a_{\;i}(u)$ is a vielbein for $\bar{g}_{ij}(u)$. 
Demanding that this frame be parallel propagated along the null geodesic 
congruence, $\bar{\nabla}_u \bar{E}^{\mu}_A=0$, imposes the condition 
\be
\del_u\bar{E}_a^i + \trac{1}{2} \bar{g}^{ij}\del_u \bar{g}_{jk}\bar{E}_a^k
= 0 \;\;\;\;\LRa\;\;\;\; 
\dot{\bar{E}}_{ai}\bar{E}^i_{\;b} = \dot{\bar{E}}_{bi}\bar{E}^i_{\;a}\;\;,
\ee
which is thus the geometric significance of the symmetry condition 
appearing in the transformation from Rosen to Brinkmann coordinates.

Now we consider geodesics $x^{\mu}(s)$ emanating from $\gamma$, 
i.e.\ $(u(0),v(0),y^i(0))=(u_0,0,0)$, with the further initial
condition that $x^{\mu\prime}(s=0)$ have no component tangent to $\gamma$,
i.e.\ vanishing scalar product with $E_-$,
\be
0=\bar{g}_{\mu\nu}(u_0)x^{\mu\prime}(0)\bar{E}_-^{\nu}(u_0) = u^{\prime}(0)\;\;.
\ee
Then the Euler-Lagrange equations following from
\be
L=u^\prime v^\prime + \trac{1}{2}\bar{g}_{ij}y^{i\prime}y^{j\prime}
\ee
imply that 
\begin{enumerate}
\item the conserved lightcone momentum $p_v$ is zero, $p_v=u^{\prime}=0$,
so that $u(s)=u_0$;
\item the transverse coordinates $y^i(s)$ evolve linearly with $s$, $y^i(s) = 
y^{i\prime}(0)s$;
\item the solution for $v(s)$ is 
$v(s) = v^{\prime}(0)s +\trac{1}{4}
\dot{\bar{g}}_{ij}(u_0)y^{i\prime}(0)y^{j\prime}(0)s^2$.
\end{enumerate}
One now introduces the geodesic coordinates $(x^{\bar{a}})=(x^{-},x^a)$
by the condition that the geodesics be straight lines, i.e.\ via
\be
x^{\bar{a}}= \bar{E}^{\bar{a}}_{\mu}x^{\mu\prime}(0)s\;\;.
\ee
Substituting this into the above solution of the geodesic equations
one finds
\be 
y^i(s) = \bar{E}^i_a x^a\;\;\;\;,\;\;\;\;
v (s) = \xm + \trac{1}{4}\dot{\bar{g}}_{ij}\bar{E}^i_a \bar{E}^j_{b}x^a x^b
\;\;,
\label{rcbc2}
\ee
which, together with $u=x^+$, is precisely the coordinate transformation
(\ref{rcbc}) from Rosen coordinates $x^{\mu}$ to Brinkmann coordinates
$x^A$. Finally we note that, as we will explain in section 4, this
transformation can also be regarded as the covariant Taylor expansion
of the $x^{\mu}$ in the quasi-transverse variables $x^{\bar{a}}$.
Here and in the following we use the terminology that ``transverse''
refers to the variables $x^a$ and ``quasi-transverse'' to the variables
$(x^{\bar{a}})=(x^-,x^a)$.

\section{\sc Null Fermi Coordinates: General Construction}

We now come to the general construction of Fermi coordinates associated
to a null geodesic $\gamma$ in a space-time with 
Lorentzian metric $g_{\mu\nu}$. 
Along $\gamma$ we introduce a parallel transported 
pseudo-orthonormal frame $E^A_{\mu}$,
\be
ds^2|_{\gamma} = 2 E^+E^- + \d_{ab}E^a E^b\;\;,
\label{frame}
\ee
with $E_+^{\mu}=\dot{\gamma}^{\mu}$, the overdot denoting the derivative 
with respect to the affine parameter. 
As in the previous section, we now consider geodesics 
$\beta(s) = (x^{\mu}(s))$ emanating from $\gamma$, i.e.\ with
$\beta(0)=x_0\in \gamma$, that satisfy
\be
g_{\mu\nu}(x_0)x^{\mu\prime}(0)E_-^{\nu}(x_0) \equiv
x^{\mu\prime}(0)E^+_{\mu}(x_0)=0\;\;.
\label{fc1}
\ee
In comparison with the standard timelike case, we note that the double
role played by the tangent vector $E_0$ to the timelike geodesic, as
the tangent vector and as the vector to which the connecting geodesics
$\beta(s)$ should be orthogonal, is in the null case shared among the
two null vectors $E_+$ (the tangent vector) and $E_-$ (providing the
condition on $\beta(s)$).

Then the Fermi coordinates $(x^A)=(x^+,x^-,x^a)$ 
of the point $x=\beta(s)$ are defined by
\be
(x^A) = (x^+,x^{\bar{a}}=sE^{\bar{a}}_{\mu}(x_0)x^{\mu\prime}(0))
\label{fc2}
\ee
where $\gamma(x^+)=x_0$ and $\bar{a}=(-,a)$. We note that these definitions
imply that 
\be
E^{\bar{a}}_{\mu}(x_0)x^{\mu\prime}(0)=
\left.\frac{\del x^{\bar{a}}}{\del s}\right|_{s=0} =
\left.\frac{\del x^{\bar{a}}}{\del x^{\mu}}\right|_{s=0}\; x^{\mu\prime}(0)
\ee
and
\be
\left.\frac{\del x^{\mu}}{\del x^+}\right|_{\gamma}
=\dot{\gamma}^{\mu}=E^{\mu}_+
\ee
so that on $\gamma$ the Fermi coordinates are related to the original
coordinates $x^{\mu}$ by
\be
\left.\frac{\del x^A}{\del x^{\mu}}\right|_{\gamma} = E^A_{\mu}\;\;\;\;,
\;\;\;\;
\left.\frac{\del x^{\mu}}{\del x^{A}}\right|_{\gamma} = E^{\mu}_{A}\;\;.
\label{dxdx}
\ee

Thus we see that Fermi coordinates are uniquely determined by a choice of
parallel pseudo-orthonormal frame along the null geodesic $\gamma$. How 
unique is this choice? Let us first consider the case of timelike Fermi
coordinates. In this case, there is a frame $(E_0,E_k)$, $k=1,\ldots,n=d+1$,
with $E_0=\dot{\gamma}$ tangent to the timelike geodesic. Evidently,
therefore, the parallel frame is unique up to constant $SO(d+1)$ rotations
of the spatial frame $E_k$. Consequently, the spatial Fermi coordinates
$x^k$, constructed exactly as above, are unique up to these constant 
rotations. 

In the lightlike case, $SO(d+1)$ is deformed to the semi-direct
product of transverse $SO(d)$-rotations of the $E_a$ (which have the obvious
corresponding effect on the transverse Fermi coordinates $x^a$) and the 
Abelian group $\simeq \RR^d$ of null rotations about $E_+$ which acts as
\be
(E_+,E_-,E_a) \mapsto (E_+,E_- -\omega^aE_a
-\trac{1}{2}\d_{ab}\omega^a\omega^b E_+,E_a + \omega_a E_+)\;\;,
\label{nr}
\ee
where $(\omega^a) \in \RR^d$ are constant parameters. Since the corresponding
action on the relevant components $E^{\bar{a}}$ of the dual frame is
\be
(E^-,E^a) \mapsto (E^-,E^a + \omega^a E^-)\;\;,
\ee
this action of constant null rotations on the frame induces the
transformation 
\be
(x^-,x^a) \mapsto (x^-,x^a + \omega^a x^-)
\ee
of the Fermi coordinates. Thus null Fermi coordinates are unique up to
constant transverse rotations and shifts of the $x^a$ by $x^-$.
This should, in particular, be compared and contrasted with the 
ambiguity 
\be
Y^k \mapsto Y^{\prime k} = Y^{\prime k}(Y^k,V)
\ee
of Penrose coordinates (\ref{acs2}), which consists of $d$ functions of
$(d+1)$ variables rather than $d(d+1)/2$ constant parameters.

For many (in particular more advanced) purposes it is useful to rephrase
the above  construction of Fermi coordinates
in terms of the Synge world function $\sigma(x,x_0)$
\cite{synge,poisson}.  For a point $x$ in the normal convex neighbourhood of
$x_0$, i.e.\ such that there is a unique geodesic $\beta$ connecting
$x$ to $x_0$, with $\beta(0)=x_0$ and $\beta(s)=x$, $\sigma(x,x_0)$
is defined by
\be
\sigma(x,x_0) = \trac{1}{2}s \int_0^s dt\; g_{\mu\nu}(\beta(t))
x^{\mu\prime}(t) x^{\nu\prime}(t)
\ee
(this is half the geodesic distance squared between $x$ and $x_0$).
Since, up to the prefactor $s$, $\sigma(x,x_0)$ is the classical action 
corresponding to the Lagrangian $L=(1/2)g_{\mu\nu}x^{\mu\prime}x^{\nu\prime}$,
standard Hamilton-Jacobi theory implies that 
\be
\sigma_{\mu}(x,x_0) \equiv \frac{\del}{\del x_0^{\mu}}\sigma(x,x_0) = 
- s g_{\mu\nu}(x_0) x^{\nu\prime}(0)\;\;,
\ee
as well as
\be
\sigma(x,x_0) = \trac{1}{2}g_{\mu\nu}(x_0)
\sigma^{\mu}(x,x_0)\sigma^{\nu}(x,x_0)\;\;.
\label{hj}
\ee
In particular, this way of writing things makes it more transparent that
something as innocuous looking as $x^{\mu\prime}(0)$ is actually a bitensor, 
namely not just a vector at $x_0$ but also a scalar at $x$.

Thus we can also summarise the construction (\ref{fc1},\ref{fc2})
of Fermi coordinates in the following way: given $x_0\in \gamma$,
the condition 
\be
\sigma^{\mu}(x,x_0)E_{\mu}^+(x_0)=0
\ee
selects those points $x$ that can be connected to $x_0$ by a geodesic with no
initial component along $\gamma$. Locally around $\gamma$ this foliates the
space-time into hypersurfaces $\Sigma_{x_0}$ pseudo-orthogonal to $\gamma$.
For $x\in \Sigma_{x_0}$, its quasi-transverse Fermi coordinates $x^{\bar{a}}$
are then defined by 
\be
x^{\bar{a}}= - \sigma^{\mu}(x,x_0) E^{\bar{a}}_{\mu}(x_0)\;\;.
\ee
Conversely, for $x\in \Sigma_{x_0}$, the $\sigma^{\mu}(x,x_0)$ can be
expressed in terms of the Fermi coordinates of $x$
(using $E^{\mu}_AE^A_{\nu}=\d^{\mu}_{\;\nu}$) as
\be
\sigma^{\mu}(x,x_0)=E^{\mu}_A
E^A_{\nu}\sigma^{\nu}(x,x_0)= - E^{\mu}_{\bar{a}}x^{\bar{a}}\;\;.
\label{sf}
\ee
It now follows from the Hamilton-Jacobi equation
(\ref{hj}) that the geodesic distance squared
of a point $x=(x^+,x^-,x^a)$ to $x_0=(x^+,0,0)$ is 
\be
2\sigma(x,x_0) = \sigma^{\mu}(x,x_0) 
E_{\mu}^A (x_0) E^{\nu}_A(x_0) 
\sigma_{\nu}(x,x_0) = \d_{ab}x^a x^b\;\;.
\ee

The $\sigma^{\mu}=\sigma^{\mu}(x,x_0)$ also appear naturally in the manifestly
covariant Taylor expansion of a function $f(x)$ around $x_0$,
\be
f(x) = 
\sum_{n=0}^{\infty}(-1)^n \frac{1}{n!}\left(\sigma^{\mu_1}\ldots\sigma^{\mu_n}
\nabla_{\mu_1}\ldots\nabla_{\mu_n}f\right)(x_0)\;\;.
\label{taylor}
\ee
This can e.g.\ be seen by beginning with the ordinary Taylor expansion
of $f(x)= f(\beta(s))$, regarded as a function of the single variable
$s$, around $s=0$, and using the geodesic equation to convert resulting
second derivatives of $x^{\mu}(s)$ into first derivatives. There is an
analogous covariant Taylor expansion for higher-rank tensor fields
\cite{poisson} which, in addition to the above component-wise covariant
expansion, also involves parallel transport from $x_0$ to $x$.

If we want to expand $f$ not around a point $x_0$ but only in the
directions quasi-transverse to a geodesic $\gamma$ with $\gamma(x^+)=x_0$,
we can use the parallel frame to project out the direction tangential to
$\gamma$. Indeed, for $x\in \Sigma_{x_0}$ 
we can use (\ref{sf}) to express $\sigma^{\mu}$ in terms of the 
quasi-transverse
Fermi coordinates $x^{\bar{a}}$. Plugging this into (\ref{taylor}),
one obtains
\be
f(x) = \sum_{n=0}^{\infty}\frac{1}{n!}
\left(E^{\mu_1}_{\bar{a}_1}\ldots E^{\mu_n}_{\bar{a}_n}
\nabla_{\mu_1}\ldots\nabla_{\mu_n}f\right)(x^+)\;x^{\bar{a}_1}\ldots
 x^{\bar{a}_n} \;\;.
\label{taylor2}
\ee
This is a Taylor expansion in the quasi-transverse Fermi coordinates
$(x^{\bar{a}})=(x^-,x^a)$, with the full dependence on $x^+$ retained.

When $f(x)$ is itself a coordinate function, $f(x)=x^{\mu}$, say,
then $\nabla_{\mu_1}f=\d_{\mu_1}^{\;\mu}$ and, for $n \geq 2$,
\be
\nabla_{(\mu_1}\ldots\nabla_{\mu_n)}f = 
- \nabla_{(\mu_1}\ldots\nabla_{\mu_{n-2}}\Gamma^{\mu}{}_{\mu_{n-1}\mu_n)}
\equiv -\Gamma^{\mu}{}_{(\mu_1\ldots\mu_n)}
\ee
(the covariant derivatives act only on the lower indices) are the generalised
Christoffel symbols. Provided that $\{x^{\mu}\}$ is an adapted coordinate
system, in the sense that $\gamma$ coincides with one of its coordinate
lines (Penrose coordinates (\ref{acs2}) are a special case of this), 
this gives us on the nose the coordinate transformation between
such adapted coordinates and Fermi coordinates, 
\be
x^{\mu}(x^+,x^{\bar{a}}) = x^{\mu}(x^+) +
E^{\mu}_{\bar{a}_1}(x^+)x^{\bar{a}_1} - 
\sum_{n=2}^{\infty}\left(\Gamma^{\mu}{}_{(\mu_1\ldots\mu_n)}
E^{\mu_1}_{\bar{a}_1}\ldots E^{\mu_n}_{\bar{a}_n}\right)(x^+)\;
x^{\bar{a}_1}\ldots x^{\bar{a}_n}\;\;.
\label{cte}
\ee
Thus the coordinate transformation between adapted and Fermi coordinates
is nothing other than the quasi-transverse Taylor expansion of the adapted
coordinates.

While formally the above equation is correct for an arbitrary coordinate
system, it is less explicit if the coordinate system is not adapted
since $x^+$, the coordinate along the geodesic, is then non-trivially 
related to the $x^{\mu}$.

In the special case of Rosen coordinates for plane waves, the above expansion
is finite and reduces to the standard result (\ref{rcbc},\ref{rcbc2}).
To see this e.g.\ for the Rosen coordinate $v$, one calculates
\be
v(x^+,x^-,x^a) = v(x^+) +
\left(\bar{E}_{\bar{a}}^{\mu}\del_{\mu}v\right)(x^+)\;x^{\bar{a}}
+\frac{1}{2}\left(\bar{E}_{\bar{a}}^{\mu}
\bar{E}_{\bar{b}}^{\nu}\nabla_{\mu}\del_{\nu}v\right)(x^+)\;
x^{\bar{a}}x^{\bar{b}}
\label{vtaylor}
\ee
with all higher order terms vanishing, and uses that on the geodesic 
$v=0$, that $\bar{E}_{-}^v=1$, $\bar{E}_a^v=0$ (\ref{rcpof}), and that
the only non-trivial $\Gamma^{v}_{\mu\nu}$
is $\Gamma^{v}_{ij}= -\frac{1}{2}\dot{\bar{g}}_{ij}$, to find yet again
\be
v = x^- + \trac{1}{4}\dot{\bar{g}}_{ij}\bar{E}^i_a \bar{E}^j_b x^a x^b
\;\;.
\ee

\section{\sc Expansion of the Metric in Null Fermi Coordinates}

We will now discuss the metric in Fermi coordinates, given by an expansion
in the quasi-transverse Fermi coordinates $x^{\bar{a}}$.  

First of all it follows from (\ref{frame}) and (\ref{dxdx}) that
to zero'th order, i.e.\ restricted to the null geodesic $\gamma$ at
$x^{\bar{a}}=0$, the metric is the flat metric.

Moreover, there are no linear terms in the metric, i.e. the Christoffel
symbols restricted to $\gamma$ are zero (the main characteristic of
Fermi coordinates in general). To see this, note that the geodesic 
equation applied to the geodesic straight lines 
\be
(x^A(s))= (x^+,x^{\bar{a}}(s)=v^{\bar{a}}s) 
\ee
implies
\be
\frac{d^2}{ds^2} x^A(s) + \Gamma^{A}_{\phantom{A}BC}\frac{d}{ds}x^{B}(s)
\frac{d}{ds}x^{C}(s) =0 \;\;\;\;\Ra\;\;\;\;
\Gamma^{A}_{\phantom{A}\bar{b}\bar{c}}(x^+,v^{\bar{a}}s)
\;v^{\bar{b}}v^{\bar{c}}=0\;\;.
\ee
Since at $s=0$ this has to be true for all $v^{\bar{a}}$, we conclude that
\be
\Gamma^A_{\phantom{A}\bar{b}\bar{c}}|_{\gamma}=0\;\;. 
\label{Gabg}
\ee
Moreover, since the 
frames $E^A_{\mu}$ are parallel propagated along $\gamma$, it
follows that in Fermi coordinates
\be
\nabla_+ E^{A}_{\mu=B}= \nabla_+ \delta^{A}_{\mu=B}= 
0 \;\;\;\;\Ra \;\;\;\; 
\Gamma^{A}_{\phantom{A}B+}|_{\gamma}=0
\;\;.
\ee
Together, these two results imply that all Christoffel symbols are zero 
along $\gamma$, 
\be
\Gamma^{A}_{\phantom{A}BC}|_{\gamma}=0\;\;.
\label{Gg}
\ee
To determine the quadratic term in the expansion of the metric, we need to
look at the derivatives of the Christoffel symbols. Differentiating 
(\ref{Gg}) along $\gamma$ one finds
\be
\Gamma^{A}_{\phantom{A}BC,+}|_{\gamma}=0\;\;.
\label{Gg+}
\ee
From the definition of the Riemann tensor
\be
R^{A}_{\phantom{A}BCD} = \Gamma^{A}_{\phantom{A}BD,C}
-\Gamma^{A}_{\phantom{A}BC,D} +
\Gamma^{A}_{\phantom{A}CE}\Gamma^{E}_{\phantom{A}BD}
-\Gamma^{A}_{\phantom{A}DE}\Gamma^{E}_{\phantom{A}BC}
\ee
it now follows that
\be
\Gamma^{A}_{\phantom{A}B+,C}|_{\gamma}= R^{A}_{\phantom{A}BC+}|_{\gamma}\;\;.
\ee
To calculate the derivatives $\Gamma^A_{\phantom{A}\bar{b}\bar{c},\bar{d}}$,
we now use the fact all the symmetrised first derivatives of the Christoffel
symbols are zero, 
\be
\Gamma^A_{\phantom{A}(\bar{b}\bar{c},\bar{d})}|_{\gamma}=0\;\;.
\label{symg}
\ee
This follows e.g.\ from applying the Taylor expansion (\ref{cte}) for
adapted coordinates to the Fermi coordinates themselves: all higher order
terms in that expansion, whose coefficients are the above symmetrised
derivatives of the Christoffel symbols, have to vanish. Incidentally, 
the required vanishing of the quadratic terms in the expansion (\ref{cte}) 
provides another argument for the vanishing (\ref{Gabg}) of the
$\Gamma^{A}_{\phantom{A}\bar{b}\bar{c}}|_{\gamma}$.

We can now calculate (with hindsight)
\be
(R^A_{\phantom{A}\bar{b}\bar{c}\bar{d}}
+R^A_{\phantom{A}\bar{c}\bar{b}\bar{d}})|_{\gamma} = 
(\Gamma^A_{\phantom{A}\bar{b}\bar{d},\bar{c}}
-\Gamma^A_{\phantom{A}\bar{b}\bar{c},\bar{d}}
+ \Gamma^A_{\phantom{A}\bar{c}\bar{d},\bar{b}}
-\Gamma^A_{\phantom{A}\bar{c}\bar{b},\bar{d}})|_{\gamma}
\ee
and use (\ref{symg}) to conclude that
\be
\Gamma^A_{\phantom{A}\bar{b}\bar{c},\bar{d}}|_{\gamma} = -\trac{1}{3}
(R^A_{\phantom{A}\bar{b}\bar{c}\bar{d}} 
+R^A_{\phantom{A}\;\bar{c}\bar{b}\bar{d}})|_{\gamma}
\;\;.
\ee 
Since we now have all the derivatives of the Christoffel symbols on
$\gamma$, we equivalently know all the second derivatives 
$g_{AB,CD}|_{\gamma}$ of the metric, namely
\bea
g_{AB,C+}|_{\gamma} &=& 0 \non
g_{++,\bar{c}\bar{d}}|_{\gamma} &=& 2 R_{+\bar{c}\bar{d}+}|_{\gamma} \non
g_{+\bar{b},\bar{c}\bar{d}}|_{\gamma} &=& -\trac{2}{3} 
(R_{+\bar{c}\bar{b}\bar{d}} +R_{+\bar{d}\bar{b}\bar{c}})|_{\gamma} \non
g_{\bar{a}\bar{b},\bar{c}\bar{d}}|_{\gamma} &=& -\trac{1}{3} 
(R_{\bar{c}\bar{a}\bar{d}\bar{b}}
+R_{\bar{c}\bar{b}\bar{d}\bar{a}})|_{\gamma}\;\;.
\eea
Thus the expansion of the metric to quadratic order is
\bea
ds^2&=& 2dx^+dx^- +\delta_{ab} dx^a dx^b \non
&-&\left[R_{+\bar{a}+{\bar{b}}} \ x^{\bar{a}} x^{\bar{b}} (dx^+)^2 
         +\trac43 R_{+{\bar{b}}\bar{a}{\bar{c}}} 
          x^{\bar{b}} x^{\bar{c}} (dx^+ dx^{\bar{a}}) +\trac13
          R_{\bar{a}{\bar{c}}{\bar{b}}{\bar{d}}}  x^{\bar{c}} x^{\bar{d}}
          (dx^{\bar{a}} dx^{\bar{b}})\right]\non
   &+& \mathcal{O}(x^{\bar{a}}x^{\bar{b}}x^{\bar{c}})
\label{metex}
\eea
where all the curvature components are evaluated on the null geodesic.
This is the precise null analogue of the Manasse-Misner result
\cite{mami,poisson} in the timelike case, i.e.\ Fermi coordinates
associated to a timelike geodesic.

In the timelike case, the expansion of the metric to fourth order was
determined in \cite{li}. The calculations in \cite{li}, based on repeated
differentiation and expansion of the geodesic and geodesic deviation
equations associated to $\gamma(u)$ and $\beta(s)$ and expressing the
results in terms of components of the Riemann tensor and its covariant
derivatives, are straightforward in principle but somewhat tedious in
practice. They can be simplified a bit by using, as we have done above,
the symmetrised derivative identities following from (\ref{cte}) instead
of the geodesic deviation equations. Either way, some care is required in
translating and adapting the intermediate steps in these calculations to
the null case (cf.\ the comment in appendix A.1). However, as far as we
can tell (and we have performed numerous checks), the final results for
the expansion of the metric in the timelike and null case are just related
by the simple index relabelling $(0,k)\lra (+,\bar{a})$, where $(x^0,x^k)$
are the Fermi coordinates in the timelike case, with $x^0$ proper time
along the timelike geodesic. In its full glory, the expansion to quartic
order (which we will require later on) is given in appendix A.1.

\section{\sc Covariant Penrose Limit Expansion via Fermi Coordinates}

We now come to the heart of the matter, namely the description of the Penrose
limit in Fermi coordinates. Let us first investigate how Fermi coordinates
transform under scalings of the metric. Thus we consider the scaling
\be
g_{\mu\nu} \ra g_{\mu\nu}(\lambda) = \lambda^{-2}g_{\mu\nu}\;\;.
\ee
First of all we note that $\gamma$ continues to be a null geodesic for the
rescaled metric. The scaling of the metric evidently
requires a concomitant scaling of the parallel pseudo-orthonormal frame
along $\gamma$, $E^A \ra E^A(\lambda)$, which must be such that
\be
2\lambda^{-2}E^+E^- + \lambda^{-2}\d_{ab}E^a E^b 
= 2E^+(\lambda)E^-(\lambda) + \d_{ab}E^a(\lambda) E^b(\lambda) 
\;\;.
\ee
Consequently, for the transverse components $E^a(\lambda)$ 
we have (up to rotations)
\be
E^a(\lambda) = \lambda^{-1}E^a\;\;.
\ee
In order to determine
the transformation of the $E^{\pm}(\lambda)$, we recall 
that in the construction of the Fermi coordinates the component $E_+$ is
fixed to be the tangent vector to $\gamma$, independently of the metric,
$E_+^{\mu}=\dot{\gamma}^{\mu}$. This requirement determines uniquely
\be
E^+(\lambda) = E^+\;\;\;\;,\;\;\;\;
E^-(\lambda) = \lambda^{-2}E^-\;\;,
\ee
which is related by a boost to the symmetric choice $E^\pm(\lambda) =
\lambda^{-1}E^\pm$. To determine the Fermi coordinates, we note that
\be
\sigma^{\mu}(x,x_0)=\trac{1}{2}sg^{\mu\nu}(x_0) 
\frac{\del}{\del x_0^{\nu}}
\int_0^s dt\; g_{\rho\sigma}(\beta(t)) 
x^{\rho\prime}(t) x^{\sigma\prime}(t)=-sx^{\mu\prime}(0)
\ee
is scale invariant. Thus the Fermi coordinates $x^A(\lambda)$ are
\bea
x^+(\lambda) &=& x^+\non 
x^-(\lambda) &=& -\sigma^{\mu}E_{\mu}^-(\lambda) = \la^{-2} x^-\non 
x^a(\lambda) &=& -\sigma^{\mu}E_{\mu}^a(\lambda) = \la^{-1} x^a \;\;.
\eea
Writing this as
\be
(x^+,x^-,x^a) = (x^+(\lambda), \lambda^2 x^-(\lambda), \lambda
x^a(\lambda))\;\;,
\label{xl}
\ee
we see that here the asymmetric rescaling of the coordinates, which is
completely analogous to that imposed ``by hand'' in Penrose 
coordinates\footnote{Here we have explicitly indicated the $\la$-dependence 
of the new coordinates that we suppressed for notational simplicity in 
(\ref{dsdsl}).},
\be
(U,V,Y^{k})=(u,\lambda^2 v(\lambda),\lambda y^{k}(\lambda))
\ee
arises naturally and automatically from the very definition of Fermi
coordinates.

To now implement the Penrose limit, 
\begin{itemize}
\item one can 
either start with the expansion (\ref{metex},\ref{ametex}) 
of the unscaled metric
in its Fermi coordinates, multiply by $\lambda^{-2}$ and express the metric
in terms of the scaled Fermi coordinates, i.e.\ make the substitution
(\ref{xl});
\item or one takes the expansion of the rescaled metric in its Fermi
coordinates $x^A(\lambda)$ and then replaces in that expansion
each $x^A(\lambda)$ by the original $x^A$.
\end{itemize}
Which point of view one prefers is a matter of taste and depends on
whether one thinks of the scale transformation actively, as acting on
space-time, or passively on measuring rods. The net effect is the same.

Let us now look at the effect of this operation on the metric
(\ref{metex},\ref{ametex}), using the language appropriate to the first point
of view to determine the powers of $\lambda$ with which each term
in (\ref{ametex}) appears. There is thus an overall $\la^{-2}$, and 
each $x^a$ or $dx^a$ contributes a $\la$ whereas $x^-$ and $dx^-$ 
gives a $\la^2$ contribution.\footnote{Alternatively, for the counting from the
second point of view, one uses the fact that the coordinate components
$\mathcal{R}(g)_{\alpha_1\cdots\alpha_n\alpha\beta}$ of the ``vertices''
$\mathcal{R}(g)_{\bar{a}_1\cdots\bar{a}_nAB}x^{\bar{a}_1}\ldots
x^{\bar{a}_n}$ appearing in the expansion of the metric $g_{AB}dx^Adx^B$
scale like the metric, $\mathcal{R}(g(\lambda))=\la^{-2}\mathcal{R}(g)$.
This can be checked explicitly for the terms written in (\ref{ametex})
and in general follows from the fact that the expansion of the metric
$g_{\mu\nu}(\lambda)$ in its Fermi coordinates $x^A(\lambda)$ must
be $\lambda^{-2}$ times the expansion of $g_{\mu\nu}$ in its Fermi
coordinates $x^A$.}
The first consequence of this is that the flat metric is of order $\lambda^0$,
the overall $\lambda^{-2}$ being cancelled
by a $\lambda^2$ from either one $dx^-$ or two $dx^a$'s. 
Moreover, precisely one of the quadratic
terms in (\ref{metex}) also gives a contribution of order $\lambda^0$,
namely $R_{a+b+}x^{a} x^{b} (dx^+)^2$, the $\lambda^{-2}$ 
being cancelled by the quadratic term in the $x^a$'s.
Thus the metric to order $\lambda^0$ is 
\be
ds^2_{\lambda^0} = 2dx^+dx^- + \d_{ab}dx^a dx^b -R_{a+b+}x^ax^b (dx^+)^2
\;\;.
\label{fpl}
\ee
Comparison with (\ref{bc}, \ref{main}) or (\ref{bc2})
shows that this is precisely the
Penrose limit along $\gamma$ of the original metric, 
\be
ds^2_{\lambda^0} = d\bar{s}^2_{\gamma}\;\;\;\;\;\;\;\;
\mathrm{(Penrose\;Limit)}
\ee
obtained here directly in Brinkmann coordinates.

Moreover the expansion to quartic order in (\ref{ametex}) is sufficient 
to give us the covariant expansion of the metric
around its Penrose limit to order
$\lambda^2$ (a quintic term would scale at least as $\la^{-2}\la^5=\la^3$).
Explicitly, the $\mathcal{O}(\lambda)$ term is
\be
ds^2_{\lambda^1} = 
  -2 R_{+a+-} \ x^a x^-
(dx^+)^2
-\trac43 R_{+bac} \ x^b x^c (dx^+ dx^a) 
-\trac13 R_{a+b+;c} \ x^a x^b x^c (dx^+)^2 
\label{plol}
\ee
and the expansion to $\mathcal{O}(\lambda^2)$ is given in appendix A.2.

One characteristic property of the lowest order (Penrose limit)
metric is the existence of the covariantly constant null vector $\del_-
\equiv \del/\del x^-$. We see from the above that $\del_-$ continues
to be null at $\mathcal{O}(\lambda)$. Actually this property is 
guaranteed to persist up to and including $\mathcal{O}(\lambda^3)$,
since a $(dx^-)^2$-term in the metric will scale at least with a power
$\la^{-2}\la^2\la^{4}=\la^4$ (such a term arises e.g.\ from the last
term in (\ref{metex}) with $\bar{a}=\bar{b}=-$ and $\bar{c}=c,\bar{d}=d$).

Moreover, we see that $\del_-$ remains Killing to $\mathcal{O}(\lambda)$
provided that $R_{+a+-}=0$. If that condition is satisfied, actually 
something more is true. Namely $\del_-$ remains covariantly constant and
the metric is that of a pp-wave (plane-fronted wave with parallel rays),
whose general form is
\be
ds^2_{\mathrm{pp}} = 2dx^+dx^- + \d_{ab}dx^a dx^b + A(x^+,x^a) (dx^+)^2 + 
2 B_b(x^+,x^a) (dx^+ dx^b)\;\;.
\ee
As shown in \cite{hs}, this is precisely the condition for string theory
in a curved background to admit a standard (conformal gauge for the
world-sheet metric $h_{rs}$) light-cone gauge $X^+(\sigma,\tau)=p_-\tau$.

More interestingly, perhaps, in general the metric to $\mathcal{O}(\lambda)$
is precisely such that it admits a modified light cone gauge $h^{00}=-1$
and $X^+(\sigma,\tau)=p_-\tau$ \cite{rudd}. 
Indeed, the conditions on the
metric $g_{AB}$ (we do not consider the conditions on the dilaton) found 
in \cite{rudd} in order for $X^-$ to have an explicit representation on 
the transverse Fock space 
\be
g_{-+}=1\;\;\;\;,\;\;\;\;
g_{-\bar{a}}=0\;\;\;\;,\;\;\;\;
\del_-^2g_{AB}=0\;\;,
\ee
(see \cite{rmat} for a discussion of the case $g_{-+}\neq 1$), 
and for $X^-$ to be auxiliary, $g_{-\bar{a}}=0$, are satisfied by the 
$\mathcal{O}(\lambda)$  metric (\ref{fpl}, \ref{plol}).

\section{\sc Example: $\mathrm{AdS}_{5}\times S^5$}

We will now illustrate the formalism introduced above by giving a
simple purely algebraic derivation of the Penrose limit expansion of the
$\mathrm{AdS}_{5}\times S^5$ metric to $\mathcal{O}(\lambda^2)$.  These
terms have been calculated before in different ways \cite{ryzhov,callan}.
In the present framework, the identification of these corrections with
certain components of the curvature tensor of $\mathrm{AdS}_{5}\times S^5$
is manifest.

Thus consider the unit (curvature) radius metric\footnote{We can restrict
to unit radius since we have already implemented the large volume limit
via the $\lambda$-expansion.}  of that space-time,
a null geodesic $\gamma$, with $E_\pm$ the lightcone components
of the corresponding parallel frame. Let us consider the case that
$\gamma$ has a non-vanishing component along the sphere (i.e.\ non-zero
angular momentum). Then, due to the product structure of the metric,
the components of $E_+$ along $S^5$ and $\mathrm{AdS}_5$ are geodesic,
and since $E_+$ is null they are of opposite norm squared $\alpha^2$.
Thus we have the decomposition
\be
E_\pm=\trac{1}{\sqrt{2}}\alpha^{\pm 1}(E_9 \pm E_0)
\ee
where $E_0$ and $E_9$ are normalised and geodesic in $\mathrm{AdS}_{5}$
and $S^5$ respectively. Without loss of generality we can (and will)
assume $\alpha=1$ because we can either perform a boost now or the
coordinate transformation $x^{\pm}\ra \alpha^{\pm 1}x^{\pm}$ later to
achieve this. We now extend $E_0$ and $E_9$ to parallel orthonormal 
frames along $\gamma$ in $\mathrm{AdS}_{5}$ and $S^5$,
\bea
ds^2_{\mathrm{AdS}} &=& \eta_{\tilde{A}\tilde{B}}E^{\tilde{A}}E^{\tilde{B}}
=-(E^0)^2 +
\d_{\tilde{a}\tilde{b}}E^{\tilde{a}}E^{\tilde{b}}\non
ds^2_{S} &=& \d_{AB}E^A E^B = \phantom{-} (E^9)^2 + \d_{ab}E^{a}E^{b}\;\;.
\eea
Here $\tilde{A},\tilde{B},\ldots =0,\ldots,4$, while 
$a,b,\ldots = 5,\ldots,8$ etc.
Since both spaces are maximally symmetric, the frame components of 
the curvature tensor are
\be
R_{\tilde{A}\tilde{B}\tilde{C}\tilde{D}}=-(
\eta_{\tilde{A}\tilde{C}} \eta_{\tilde{B}\tilde{D}}
-\eta_{\tilde{A}\tilde{D}} \eta_{\tilde{B}\tilde{C}})\;\;\;\;,\;\;\;\;
R_{ABCD} = \d_{AC}\d_{BD}-\d_{AD}\d_{BC}\;\;,
\ee
and therefore the only non-vanishing frame components in the parallel frame
$(E_\pm,E_{\tilde{a}},E_a)$ along $\gamma$ are
\bea
&&R_{\tilde{a}\tilde{b}\tilde{c}\tilde{d}} = -(
\d_{\tilde{a}\tilde{c}}\d_{\tilde{b}\tilde{d}}-
\d_{\tilde{a}\tilde{d}}\d_{\tilde{b}\tilde{c}})\;\;\;\;\;\;\;\;\;\;\;
R_{abcd} =\d_{ac}\d_{bd}-\d_{ad}\d_{bc}\non
&&R_{+a+b}=R_{-a-b}=\phantom{-}R_{+a-b}=\phantom{-}R_{-a+b} 
= \trac{1}{2}\d_{ab}\non
&&R_{+\tilde{a}+\tilde{b}}=R_{-\tilde{a}-\tilde{b}}=
-R_{+\tilde{a}-\tilde{b}}=-R_{-\tilde{a}+\tilde{b}} =
\trac{1}{2}\d_{\tilde{a}\tilde{b}}
\label{rpof}
\eea
We now have all the information we need to determine the Penrose limit 
and the higher order corrections. For the Penrose limit we immediately
find, from (\ref{fpl}), the result\footnote{Here and in the following 
we use a short-hand notation, 
$\tilde{x}^2 = \d_{\tilde{a}\tilde{b}}x^{\tilde{a}} x^{\tilde{b}}$, 
$xdx = \d_{ab}x^a dx^b$, etc.}
\be
ds^2_{\lambda^0}= 2dx^+dx^- + dx^2 + d\tilde{x}^2  
-\trac{1}{2}(x^2 + \tilde{x}^2)(dx^+)^2\;\;.
\ee
This is of course the standard result \cite{bfhp2,bmn}, namely the
maximally supersymmetric BFHP plane wave \cite{bfhp1}.

On symmetry grounds and/or because the curvature tensors are covariantly
constant, all the $\mathcal{O}(\lambda)$-corrections (\ref{plol}) to the 
Penrose limit are identically zero in this case. Actually, (\ref{rpof})
shows that to any order only even numbers of transverse indices
$(a,b,\ldots)$ or $(\tilde{a},\tilde{b},\ldots)$ can appear in 
the expansion of the metric, and thus all odd order corrections
$\mathcal{O}(\la^{2n+1})$ to the metric are identically zero.

For the
$\mathcal{O}(\lambda^2)$-corrections, displayed in (\ref{ploll}), 
one finds non-zero contributions from the second, fourth and fifth terms in
square brackets as well as from the term quadratic in the Riemann tensor,
and one can read off the result
\bea
ds^2 &=& 2dx^+dx^- + dx^2 + d\tilde{x}^2 -
(x^2+\tilde{x}^2)(dx^+)^2\non
     &+&\lambda^2\left[ -\trac{2}{3}(x^2-\tilde{x}^2) (dx^+dx^-) 
-\trac{1}{3}(x^2 dx^2 -(xdx)^2)
 + \trac{1}{3} (\tilde{x}^2 d\tilde{x}^2 - 
(\tilde{x}d\tilde{x})^2)\right.\non
&& \phantom{\la^2\;\,}\left. +\trac{2}{3} x^-(xdx-\tilde{x}d\tilde{x}) dx^+ 
+ \trac{1}{6} ((x^2)^2-(\tilde{x}^2)^2) (dx^+)^2\right] +
\mathcal{O}(\lambda^4)
\eea
While this may not be the world's nicest metric, at least every term in this
metric has a clear geometric interpretation in terms of the Riemann tensor of
the original $\mathrm{AdS}\times S$ metric. This metric can be simplified
somewhat, perhaps at the expense of geometric clarity, by the 
$\lambda$-dependent coordinate transformation
\be
x^-=w^-(1-\frac{\lambda^2}{6}(y^2-z^2))\;\;,\;\;
x^a= y^a(1-\frac{\la^2}{12}y^2)\;\;,\;\;
x^{\tilde{a}}= z^a(1+\frac{\la^2}{12}z^2)\;\;,
\label{cct}
\ee
which has the effect of removing the explicit $x^-$ from the metric and
eliminating the radial $xdx$ and $\tilde{x}d\tilde{x}$ terms. Performing
only the $x^-$-transformation, and neglecting
terms of $\mathcal{O}(\la^4)$, the metric takes the form 
\bea
ds^2 &=& 2dx^+dw^- + dx^2 + d\tilde{x}^2 -
(x^2+\tilde{x}^2)(dx^+)^2\non
     &+&\frac{\lambda^2}{3}\left[ -3(x^2-\tilde{x}^2) (dx^+dw^-) 
-(x^2 dx^2 -(xdx)^2)
 + (\tilde{x}^2 d\tilde{x}^2 - 
(\tilde{x}d\tilde{x})^2)\right.\non
&& \phantom{\la^2\;\,}\left. 
+ ((x^2)^2-(\tilde{x}^2)^2) (dx^+)^2\right] +
\mathcal{O}(\lambda^4)\;\;.
\eea
With $w^- \ra -2x^-$ and  $\lambda \ra 1/R$, $R$ the radius, this agrees with
the metric found in \cite{ryzhov}. The subsequent
transformation $(x^a,x^{\tilde{a}})\ra (y^a,z^a)$ leads to the metric
\bea
ds^2 &=& 2dx^+dw^- + dy^2 + dz^2 - (y^2 + z^2) (dx^+)^2\non
&&   + \frac{\lambda^2}{2}\left[(y^4-z^4)(dx^+)^2 - 2 (y^2-z^2)dx^+dw^- +
z^2dz^2 - y^2 dy^2\right]\;\;,
\eea
which, with $w^-\ra x^-$, is identical to the metric found in
\cite{callan} (via a coordinate transformation similar to (\ref{cct})
before taking the Penrose limit) and studied there from the point of
view of the BMN correspondence \cite{bmn}.

\section{\sc A Peeling Theorem for Penrose Limits}

In section 6 we have seen that the leading non-trivial contribution to
the metric in a series expansion in the scaling parameter $\lambda$ arises
at $\mathcal{O}(\lambda^0)$ from the $R_{a+b+}$ component of the Riemann
tensor. And, more generally, we have essentially already seen (and used)
there, although we did not phrase it that way, that under a rescaling
\be
g_{\mu\nu}\ra g(\lambda)_{\mu\nu}=\la^{-2}g_{\mu\nu}
\ee
of the metric, effectively
the components $R_{ABCD}$ of the Riemann tensor restricted to the null
geodesic scale as
\be
R_{ABCD}(g(\lambda))=\la^{-2+w_A+w_B+w_C+w_D}R_{ABCD}(g)
\ee
where the weights are 
\be
(w_+,w_-,w_a) = (0,2,1)\;\;.
\label{w}
\ee
The resulting scaling weights $w=-2+w_A+w_B+w_C+w_D$ 
of the frame components of the 
Riemann tensor are summarised in the table below.

\[
\begin{array}{|c|c|c|c|c|}\hline
\lambda^0& \lambda^1& \lambda^2& \lambda^3& \lambda^4 \\ \hline
R_{a+b+} &
R_{+-+a}, R_{+abc} &
R_{+-+-}, R_{+a-b}, R_{+-ab}, R_{abcd} & 
R_{+-a-}, R_{-abc} & 
R_{-a-b}\\
\hline
\end{array}
\]

It is also not difficult to see that
the leading scaling weight of a component of the Riemann (Weyl)
tensor at a point $x$ not on $\gamma$ is identical to that on $\gamma$,
\be
R_{ABCD}(x_0) = \mathcal{O}(\la^w)\;\;\;\;\Ra\;\;\;\;
R_{ABCD}(x) = \mathcal{O}(\la^w)\;\;.
\label{rxrx}
\ee
To be specific, in this equation we let both $R_{ABCD}(x_0)$ and
$R_{ABCD}(x)$ refer to frame components at the respective points (since
the generalised Petrov classification \cite{cmpp} we will employ below
refers to such components), the frame at $x$ being obtained by parallel
transport of the standard frame at $x_0$ along the unique geodesic
connecting $x$ and $x_0$.

The statement (\ref{rxrx}) is intuitively obvious since moving away
from $\gamma$ involves more insertions of quasi-transverse coordinates
$x^{\bar{a}}$ and thus, upon scaling of the coordinates, higher powers
of $\lambda$.  One can base a formal argument along these lines on the
covariant Taylor expansion of a tensor. However, for present purposes it
is enough to note that the expansion of a tensor at a point $x=(x^+,\la^2
x^-,\la x^a)$ around the point $x_0=(x^+,0,0)$ is tantamount to an
expansion in non-negative powers of $\lambda$. The same is true for the
frames and this establishes (\ref{rxrx}).  This argument also shows that
the statement (\ref{rxrx}) as such is also valid for Fermi coordinate
rather than frame components since they agree at $x_0$ and differ by
higher powers of $\lambda$ at $x$.

We will now establish the relation of the above results to the peeling
property of the Weyl tensor in the Penrose limit context. This was first
analysed in the four-dimensional $d=2$ case in \cite{kunze}, where it
was shown that the complex Weyl scalars $\Psi_i$, $i=0,\ldots,4$ scale
as $\la^{4-i}$, the $\mathcal{O}(\lambda^0)$-term $\Psi_4$ corresponding
to the type N Penrose limit components $C_{a+b+}$.

In higher dimensions $d>2$, instead of complex
Weyl scalars (one complex transverse dimension) one has $SO(d)$-tensors
of the transverse rotation group, and the appropriate framework is then
provided by the analysis in \cite{cmpp}. There the primary classification
of the Weyl tensor (according to principal or Weyl type) is based on the
boost weight of a frame component of a tensor under the boost
\be
(E_+,E_-) \ra (\alpha^{-1} E_+, \alpha E_-) 
\ee
Evidently, the individual boost weights $b_A$ are 
\be
(b_+,b_-,b_a)=(-1,+1,0)\;\;.
\ee
Comparison with (\ref{w}) shows that 
$b_A = w_A -1$, and thus the relation between $w$ and the boost weight 
$b=\sum b_A$ of the Riemann or Weyl tensor is
\be
b= \sum_A (w_A-1) = w-2 \in \{-2,-1,0,1,2\}\;\;.
\ee
In particular, the characterisation in terms of the scaling weight $w$
is equivalent to that in terms of boost weights, and a component with
boost weight $b$ scales as $\lambda^{b+2}$.

According to the generalised Petrov classification in \cite{cmpp}, the
component characterising the alignment property of type N has the lowest
boost weight $b=-2$, thus scales as $\lambda^0$, as we already know
from the Penrose limit, type III has $b=-1$, etc.\footnote{In comparing
with \cite{cmpp}, one should note that there the metric decays along
the null geodesic (connecting an interior point to conformal infinity)
whereas here this decay occurs in the directions quasi-transverse to
the null geodesic. Thus their $C_{0i0j}$ correspond to our $C_{-a-b}$
etc.} Thus, generalising the result of \cite{kunze}, we have established
that the scaling properties (scaling weights) of the frame components of
the Weyl tensor are strictly correlated with their algebraic properties.
This can be regarded as a formal analogue, in the Penrose limit context,
of the standard peeling theorem \cite{peeling} of radiation theory in 
general relativity 
which describes the algebraic properties of the coefficients of the Weyl
tensor in a large distance $1/r$ expansion.

\subsection*{Acknowledgements}

This work has been supported by the Swiss National Science Foundation
and by the EU under contract MRTN-CT-2004-005104.

\appendix

\section{\sc Higher Order Terms}

\subsection{\sc Expansion of the Metric in Fermi Coordinates to Quartic Order}

As mentioned in section 5, the expansion of the metric in null Fermi
coordinates follows the pattern of the expansion in the timelike case, 
determined to quartic order in \cite{li}. Thus one has\footnote{In 
the second line, the Manasse-Misner result \cite{mami,poisson},
we have corrected a misprint in \cite{li}.}
\bea
ds^2&=& 2dx^+dx^- +\delta_{ab} dx^a dx^b \non
&&-R_{+\bar{a}+{\bar{b}}} \ x^{\bar{a}} x^{\bar{b}} (dx^+)^2 
         -\trac43 R_{+{\bar{b}}\bar{a}{\bar{c}}} 
          x^{\bar{b}} x^{\bar{c}} (dx^+ dx^{\bar{a}}) -\trac13
          R_{\bar{a}{\bar{c}}{\bar{b}}{\bar{d}}}  x^{\bar{c}} x^{\bar{d}}
          (dx^{\bar{a}} dx^{\bar{b}})\non
&&-\trac13 R_{+{\bar{a}}+{\bar{b}};{\bar{c}}} \ 
          x^{\bar{a}} x^{\bar{b}} x^{\bar{c}} (dx^+)^2   
          -\trac14 R_{+{\bar{b}}{\bar{a}}{\bar{c}};{\bar{d}}} \ x^{\bar{b}}
          x^{\bar{c}} x^{\bar{d}} (dx^+ dx^{\bar{a}}) 
         -\trac16 R_{{\bar{a}}{\bar{c}}{\bar{b}}{\bar{d}};{\bar{e}}} \
          x^{\bar{c}} x^{\bar{d}} x^{\bar{e}}(dx^{\bar{a}} dx^{\bar{b}})\non
&&+(\trac13 R_{+{\bar{a}}A{\bar{b}}}
  R^A_{\phantom{A}{\bar{c}}+{\bar{d}}} -\trac1{12}
  R_{+{\bar{a}}+{\bar{b}};{\bar{c}}{\bar{d}}}) \ x^{\bar{a}} x^{\bar{b}}
  x^{\bar{c}} x^{\bar{d}} (dx^+)^2  \non
     & &+ (\trac2{15} R_{+{\bar{b}}A{\bar{c}}}
       R^A_{\phantom{A}{\bar{d}}{\bar{a}}{\bar{e}}} -\trac1{15}
       R_{+{\bar{b}}{\bar{a}}{\bar{c}};{\bar{d}}{\bar{e}}})
       \ x^{\bar{b}} x^{\bar{c}} x^{\bar{d}} x^{\bar{e}}(dx^+
       dx^{\bar{a}}) \non 
   &&+(\trac2{45} R_{A{\bar{c}}{\bar{a}}{\bar{d}}}
       R^A_{\phantom{A}{\bar{e}}{\bar{b}}{\bar{f}}} -\trac1{20}
       R_{{\bar{a}}{\bar{f}}{\bar{b}}{\bar{c}};{\bar{d}}{\bar{e}}})
       \ x^{\bar{c}} x^{\bar{d}} x^{\bar{e}}x^{\bar{f}} (dx^{\bar{a}}
       dx^{\bar{b}})\non
   && + \mathcal{O}(x^{\bar{a}}x^{\bar{b}}x^{\bar{c}}x^{\bar{d}}x^{\bar{e}})
\label{ametex}
\eea
However, the actual calculation of the fourth and higher order terms 
requires a closer inspection. For example, to determine
the metric at quartic order, one needs to express the third derivatives of the
Christoffel symbols in terms of Riemann tensors. One such identity is
\bea
\Gamma^A_{++,\bar{a}\bar{b}\bar{c}}&=&
R^A_{\phantom{\mu}+(\bar{a}|+;|\bar{b}\bar{c})}+
R^A_{\phantom{\mu}(\bar{a}\bar{b}|+;|\bar{c})+}
-R^A_{\phantom{\mu}+(\bar{a}|B}R^B_{\phantom{\mu}|\bar{b}|+|\bar{c})}\non
&&
-3R^A_{\phantom{\mu}B(\bar{a}|+}R^B_{\phantom{\mu}|\bar{b}|+|\bar{c})}
+3R^A_{\phantom{\mu}(\bar{a}|B|\bar{b}|}R^B_{\phantom{\mu}+|\bar{c})+}
-2R^A_{\phantom{\mu}(\bar{a}|\bar{p}|\bar{b}|}
R^{\bar{p}}_{\phantom{\bar{p}}+|\bar{c})+}
\label{li2}
\eea
As written, this identity is correct both in the null and (with the
substitution $(+,\bar{a})\ra(0,k)$) in the timelike case, whereas the 
expression given in \cite[eq.(33)]{li},
\bea
\Gamma^\mu_{00,klm}&=&R^\mu_{\phantom{\mu}0(k|0;|lm)}+
R^\mu_{\phantom{\mu}(kl|0;|m)0}
-R^\mu_{\phantom{\mu}0(k|\kappa}R^\kappa_{\phantom{\mu}|l|0|m)}\non
&&
-3R^\mu_{\phantom{\mu}\kappa(k|0}R^\kappa_{\phantom{\mu}|l|0|m)}
+R^\mu_{\phantom{\mu}(k|p|l|}R^p_{\phantom{p}0|m)0}
\eea
is valid only in the timelike case (where it agrees with (\ref{li2})).

\subsection{\sc Expansion around the Penrose Limit to $\mathcal{O}(\la^2)$}

The covariant Fermi coordinate expansion of the Penrose limit to
$\mathcal{O}(\la^2)$ is
\bea
ds^2 &=& 2dx^+dx^- + \d_{ab}dx^a dx^b -R_{a+b+}x^ax^b (dx^+)^2\non
     &+& \lambda\left[-2 R_{+a+-} \ x^a x^-
(dx^+)^2 -\trac43 R_{+bac} \ x^b x^c (dx^+ dx^a) 
-\trac13 R_{+a+b;c} \ x^a x^b x^c (dx^+)^2 \right]\non
&+&\la^2\left[-R_{+-+-} \ x^- x^- (dx^+)^2 
          -\trac43 R_{+b-c} \ x^b x^c (dx^+ dx^-) -\trac43 R_{+-ac} \ x^-
x^c (dx^+ dx^a)\right.\non
&& -\trac43 R_{+ba-} \ x^b x^- (dx^+ dx^a)
          -\trac13 R_{acbd} \ x^c x^d (dx^a dx^b) 
 -\trac23 R_{+a+-;c} \ x^a x^- x^c (dx^+)^2\non
&& -\trac13 R_{+a+b;-} \ x^a x^b x^- (dx^+)^2  
-\trac14 R_{+bac;d} \ x^b x^c x^d (dx^+ dx^a) \non
&&\left.+(\trac13 R_{+aAb} R^A_{\;c+d} -\trac1{12} R_{+a+b;cd}) \ x^a x^b
x^c x^d (dx^+)^2\right]\non
&& + \mathcal{O}(\la^3) 
\label{ploll}
\eea
Determining the expansion to $\mathcal{O}(\la^3)$ would require knowledge
of the quintic terms in the expansion of the metric in Fermi coordinates.

\rnc{\Large}{\normalsize}

\end{document}